\documentclass[10pt]{article}

\usepackage{color} 
\usepackage{comment} 
\usepackage{amssymb}
 
\def \ccomma{\raise 2pt\hbox{,}\ } 
\def \D {\hbox{d}}

\def \varphiKR{\phi}
\def \tg {\mathop{\rm tg}\nolimits}
\def \grad{\mathop{\rm grad}\nolimits}

\def \Pn     {{\rm P_{\rm n}}}
\def \PVI    {{\rm P_{\rm VI}}}
\def \PV     {{\rm P_{\rm V}}}
\def \PIV    {{\rm P_{\rm IV}}}
\def \PIII   {{\rm P_{\rm III}}}
\def \PII    {{\rm P_{\rm II}}}
\def \PI     {{\rm P_{\rm I}}}
 
\def \Cn     {{\rm C_{\rm n}}}
\def \CVI    {{\rm C_{\rm VI}}}  
                                                
\def \CVa    {{\rm C_{\rm Va}}}  
\def \CVb    {{\rm C_{\rm Vb}}}  
\def \CIII   {{\rm C_{\rm III}}}
\def \CIV    {{\rm C_{\rm IV}}}

\def \rKtwo  {r}
\def \ka     {\lambda}
\def \cu     {c_u}
\def \cv     {c_v}
\def \du     {{d_u}} 
\def \dv     {{d_v}} 
\def \Kcinq  {{K_5}}
\def \Ksix   {{K_6}}
\def \Ksept  {{K_7}}
\def \fVI  {{\rm f_{\rm VI}}}
\def \gVI  {{\rm g_{\rm VI}}}

\def \farbone {F} 
\def \fa  {a}
\def \fb  {b}
\def \fx  {f}
\def \fy  {g}
\def \FG  {\Phi}
\def \xif  {\xi}
\def \hsum {A}
\def \hdif {B}
\def \cone{c_1}

\def\Ulin {\tilde{u}}
\def\Vlin {\tilde{v}}

\def\UU {\tilde{U}}
\def\VV {\tilde{X}}
\def\xxi{\tilde{\xi}}

\def \xiofZ  {f}
\def \psiofZ {\psi}
\def \farbx  {\fx}
\def \farby  {\fy}

\def \farbt  {h_0}
\def \Fu     {F}
\def \Fvt    {G}
\def \phiFu  {\rho} 
\def \eff    {\lambda}
\def \egg    {\mu}

\begin{document}

\title{R\'eductions d'un syst\`eme bidimensionnel de sine-Gordon \`a 
la sixi\`eme \'equation de Painlev\'e
}

\author{Robert Conte${}^{1,2}$ et A.~Michel Grundland${}^{3}$
{}\\
\\ 1. Universit\'e Paris-Saclay, ENS Paris-Saclay, CNRS,
\\    Centre Borelli, F-91190 Gif-sur-Yvette, France
\\
\\ 2. Department of mathematics, The University of Hong Kong,
\\ Pokfulam, Hong Kong.
\\
\\ 3. D\'epartement de math\'ematiques et d'informatique,
\\ Universit\'e du Qu\'ebec \`a Trois-Rivi\`eres,
   Qu\'ebec, G9A 5H7, Canada
\\
\\    Courriel Robert.Conte@CEA.FR,       \\ ORCID https://orcid.org/0000-0002-1840-5095
\\    Courriel Grundlan@CRM.UMontreal.CA, \\ ORCID https://orcid.org/0000-0003-4457-7656
{}\\
}

\maketitle

\hfill 

{\vglue -10.0 truemm}
{\vskip -10.0 truemm}

\begin{abstract}
Nous \'etablissons toutes les r\'eductions du syst\`eme de deux \'equations coupl\'ees de sine-Gordon 
introduit par Konopelchenko et Rogers
\`a des \'equations diff\'erentielles ordinaires.
Ces r\'eductions sont toutes des d\'eg\'en\'erescences 
d'une r\'eduction ma{\^\i}tresse 
\`a une \'equation jug\'ee par Chazy 
``curieuse en raison de [son] \'el\'egance'',
transform\'ee alg\'ebrique de la sixi\`eme \'equation de Painlev\'e 
la plus g\'en\'erale.

\end{abstract}
 


\noindent \textit{Mots-clefs}:
sine-Gordon bidimensionnel,
r\'eductions,
sixi\`eme \'equation de Painlev\'e.

\medskip


\baselineskip=12truept 

\tableofcontents

\section{Introduction: un syst\`eme de sine-Gordon coupl\'e}

Gauss a montr\'e que 
l'\'equation de sine-Gordon $\phi_{xt}+\sin(\phi)=0$
caract\'erise les surfaces \`a courbure constante,
et la question s'est longtemps pos\'ee d'une \'eventuelle extra\-polation
\`a une variable ind\'ependante de plus.
La premi\`ere extrapolation int\'egrable propos\'ee \cite{BLPsine-Gordon}
n'\'etait pas une \'equation aux d\'eriv\'ees partielles (EDP)
mais une \'equation int\'egro-diff\'erentielle.
Une extrapolation \`a une EDP int\'egrable 
a \'et\'e ult\'erieurement introduite par Konopelchenko et Rogers \cite[Eq.~(47)]{KR-PLA},
\begin{eqnarray}
& &
\left[e^{ i (\varphiKR+\psi)} \varphiKR_{XT}\right]_X - \sigma^2
\left[e^{ i (\varphiKR+\psi)} \varphiKR_{YT}\right]_Y =0, \sigma^4=1,
\\ \nonumber\ & &
\left[e^{-i (\varphiKR+\psi)}    \psi_{XT}\right]_X - \sigma^2
\left[e^{-i (\varphiKR+\psi)}    \psi_{YT}\right]_Y =0,
\label{eqsysKR}
\end{eqnarray}  
dont la limite $\partial_Y \to 0$ (resp.~$\partial_X \to 0$) \cite[(1.2)]{KD-SAM.2d-sG} est bien 
sine-Gordon pour $\varphiKR(X,T)$ (resp.~$\psi(Y,T)$). 

Une rotation dans les plans  $(\varphiKR,\psi)$ et $(X,Y)$ \cite[(2.1)]{KD-SAM.2d-sG}
(le param\`etre inessentiel $\nu$ sert \`a unifier les notations de divers auteurs), 
\begin{eqnarray}
& &
u=\frac{\varphiKR+\psi}{2 \nu}\ccomma
v= i \frac{\varphiKR-\psi}{2}\ccomma
x=\sigma X+Y,
y=\sigma X-Y,
t=T,
\end{eqnarray} 
conduit au syst\`eme \'equivalent
\begin{eqnarray}
& &
u_{xyt} + u_x v_{yt} + u_y v_{xt}=0,
\left(v_{xy} -\nu^2 u_x u_y\right)_t=0.
\label{eqsysuv} 
\end{eqnarray}
Ce syst\`eme, qui admet la r\'eduction $v=\pm i \nu u$,
est l'objet de la pr\'esente \'etude.
Il est invariant par un changement arbitraire des variables ind\'ependantes 
$(x,y,t) \to (\varphi_1(x),\varphi_2(y),\varphi_3(t))$,
et c'est la condition de compatibilit\'e d'un syst\`eme de trois \'equations lin\'eaires
\cite[Eq.~(12)]{KR-PLA}.
Cette condition, \'equivalente \`a \cite[Eq.~(2.5)]{KD-SAM.2d-sG},
\begin{eqnarray}
& &
\left\lbrace
\begin{array}{ll}
\displaystyle{
L_1 \Psi=0, 
L_2 \Psi=0, 
[L_1,L_2] \Psi=0,
}\\ \displaystyle{
L_1=\pmatrix{\partial_x & - \nu^2 u_x \cr u_y & \partial_y \cr},
L_2=\pmatrix{\partial_y \partial_t+v_{yt} & u_y \partial_t \cr u_x \partial_t & \partial_x \partial_t + v_{xt} \cr},
}
\end{array}
\right.
\label{eqL1L2A1A2}  
\end{eqnarray}
permet la construction explicite du $N$-soliton \cite[(4.25), (4.32)]{KD-SAM.2d-sG},
donc le syst\`eme (\ref{eqsysuv}) poss\`ede la propri\'et\'e de Painlev\'e. 

Notre motivation est la suivante.
L'\'equation de sine-Gordon,
qui caract\'erise les surfaces \`a courbure constante,
admet une r\'eduction \`a la troisi\`eme \'equation de Painlev\'e $\PIII$,
et il existe un syst\`eme \cite[Eq.~(9)]{RGH1991} 
compos\'e de deux \'equations discr\`etes de $\PIII$ 
qui s'int\`egrent par une version discr\`ete de $\PVI$ \cite[Eq.~(1.1)]{JSRG},
la plus g\'en\'erale des \'equations de Painlev\'e \cite{CMBook2}.
Le syst\`eme (\ref{eqsysuv}), qui couple deux \'equations de sine-Gordon,
pourrait donc admettre une r\'eduction \`a $\PVI$.
\medskip

Notre but est donc de trouver, si elle existe,
une r\'eduction de (\ref{eqsysuv}) \`a $\PVI$.
Remarquons que l'ordre diff\'erentiel (six) du syst\`eme (\ref{eqsysuv})
laisse en effet esp\'erer l'existence d'une r\'eduction \`a la $\PVI$ la plus g\'en\'erale,
une EDO d'ordre deux \`a quatre param\`etres.

Dans la section \ref{sectionComplet},
nous montrons d'abord l'inexistence de termes compl\'ementaires du syst\`eme (\ref{eqsysuv}).
La section \ref{sectionReductionsODE}
d\'efinit et int\`egre,
sans recourir \`a la th\'eorie des groupes,
toutes les r\'eductions possibles de l'EDP (\ref{eqsysuv}) \`a une EDO.
Enfin, dans la section \ref{sectionReductionsPDE},
nous isolons celles d'entre elles qui sont vraiment des r\'eductions.

\section{Syst\`eme complet engendr\'e par le syst\`eme d'EDPs (\ref{eqsysuv})}
\label{sectionComplet}

Puisqu'il poss\`ede la propri\'et\'e de Painlev\'e, 
le syst\`eme (\ref{eqsysuv}) r\'eussit \'evidemment au test \cite{CMM1996,RL-2dimSG}. 
\'Etablissons n\'eanmoins l'emplacement des fonctions arbitraires
(indices de Fuchs) dans les s\'eries de Laurent des champs $\grad u$, $\grad v$,
dont nous aurons besoin par la suite. 
Ces champs poss\`edent des p\^oles mobiles simples,
leurs s\'eries de Laurent sont donc d\'efinies par
\begin{eqnarray}
& & 
u=(a/ \nu) \log \varphi + \sum_{j=0}^{+\infty} u_j(x,y,t) \varphi^j,\
v=b        \log \varphi + \sum_{j=0}^{+\infty} v_j(x,y,t) \varphi^j,\ 
\label{eqLaurentuv}
\end{eqnarray} 
o\`u $\varphi(x,y,t)=0$ repr\'esente la vari\'et\'e singuli\`ere mobile.
Les termes de degr\'e de singularit\'e minimal (trois) en $\varphi$ d\'efinissent deux 
familles: $a= \pm i,b=1$.
\smallskip
L'\'equation lin\'earis\'ee au voisinage d'une des deux familles
%
%
\begin{eqnarray}
& &
\left\lbrace
\begin{array}{ll}
\displaystyle{
\nu u \sim i \log \varphi + \varepsilon \Ulin,v \sim \log\varphi+ \varepsilon \Vlin, \varepsilon\to 0,
}\\ \displaystyle{
\Ulin_{xyt}
 +i\frac{\varphi_y}{\varphi} \Vlin_{xt} 
 +i\frac{\varphi_x}{\varphi} \Vlin_{yt} 
 -\frac{\varphi_y \varphi_t}{\varphi^2} \Ulin_{x}
 -\frac{\varphi_x \varphi_t}{\varphi^2} \Ulin_{y}=0,
}\\ \displaystyle{
\Vlin_{xyt}
 -i\frac{\varphi_y}{\varphi} \Ulin_{xt} 
 -i\frac{\varphi_x}{\varphi} \Ulin_{yt} 
 +i\frac{\varphi_y \varphi_t}{\varphi^2} \Ulin_{x}
 +i\frac{\varphi_x \varphi_t}{\varphi^2} \Ulin_{y}=0,
}
\end{array}
\right.
\end{eqnarray}
admet des solutions du type de Fuchs 
\begin{eqnarray}
& & 
\Ulin= \Ulin_0 \varphi^{j}\left[1+O(\varphi)\right],
\Vlin= \Vlin_0 \varphi^{j}\left[1+O(\varphi)\right],
\end{eqnarray} 
et la condition $\Ulin_0 \Vlin_0 \not=0$ engendre l'\'equation indicielle
\begin{eqnarray}
& &
\det \pmatrix{\nu j^2 (j-3) & 2 i j (j-1)  \cr 2 \nu j (j-2) & i j (j-1)(j-2) \cr} =0,
\end{eqnarray} 
dont les six racines sont les indices de Fuchs $j=-1,0,0,1,2,4$.
Les six fonctions arbitraires associ\'ees sont respectivement
$\varphi$ et les coefficients $u_0,v_0,v_1,v_2,v_4$ de (\ref{eqLaurentuv}).
\medskip

Puisque tous les termes du syst\`eme (\ref{eqsysuv})
ont le m\^eme degr\'e de singularit\'e (trois),
c'est un syst\`eme simplifi\'e (au sens classique \cite[\S 37]{PaiBSMF}),
dont il importe de d\'eterminer les termes compl\'ementaires,
ceux dont l'ajout conserve la propri\'et\'e de Painlev\'e.
Ces termes compl\'ementaires sont
ici le polyn\^ome le plus g\'en\'eral 
des dix-huit d\'eriv\'ees premi\`eres et secondes de $u$ et de $v$ 
\`a coefficients fonctions de $(x,y,t)$ 
de degr\'e de singularit\'e au plus deux.
Ce calcul classique, que nous ne d\'etaillerons pas ici, 
et qui consiste essentiellement \`a exiger l'existence des deux s\'eries de Laurent (\ref{eqLaurentuv})
en annulant la contribution d'\'eventuels logarithmes,
s\'electionne parmi les cinquante-six termes possibles dix termes 
d\'ependant de deux fonctions arbitraires.
Mais le syst\`eme complet 
se ram\`ene alors au syst\`eme simplifi\'e par une translation de $(u,v)$,
donc le syst\`eme (\ref{eqsysuv}) n'admet aucun terme compl\'ementaire.

\section{R\'eductions possibles} 
\label{sectionReductionsODE}

Consid\'erons les r\'eductions de (\ref{eqsysuv}) \`a un syst\`eme d'EDOs
d\'efinies par
\begin{eqnarray}
& &
\left\lbrace
\begin{array}{ll}
\displaystyle{
\xi=\xif(x,y,t), 
}\\ \displaystyle{
u=\cu(x,y,t) U(\xi) + \nu^{-1} \du(x,y,t),
}\\ \displaystyle{
v=\cv(x,y,t) V(\xi) +\int \dv(x,y,t) \D t.
}
\end{array}
\right.
\label{eqred-no-UV}
\end{eqnarray} 

Celles d'entre elles qui sont caract\'eristiques,
c'est-\`a-dire qui abaissent l'ordre diff\'erentiel,
engendrent un syst\`eme lin\'eaire d'EDOs,
nous les excluons et supposons d\'esormais $\xif_x \xif_y \xif_t \not=0$. 

Toute r\'eduction non-caract\'eristique de (\ref{eqsysuv}) \`a un syst\`eme d'EDOs
a pour ordre diff\'eren\-tiel six.
Puisque la propri\'et\'e de Painlev\'e est conserv\'ee \cite[\S 4.3]{CMBook2} lors d'une telle r\'eduction
$(u,v)(x,y,t) \to (U,V)(\xi), \xi=\xi(x,y,t)$,
le syst\`eme r\'eduit ne peut d\'ependre que des d\'eriv\'ees de $U,V$,
\`a l'exclusion de $U,V$,
il est donc d'ordre diff\'erentiel quatre en $U',V'$.
Ces r\'eductions non-caract\'eristiques engendrent le syst\`eme qualitatif,
\begin{eqnarray}
& &
\left\lbrace
\begin{array}{ll}
\displaystyle{
%
U''' + 2 \cv            U' V'' + f_1 U'' + f_2  U' V' + f_4 V'' + f_3 U' + f_5 V'+ f_6=0,
}\\ \displaystyle{
V''' - 2 \cu^2 \cv^{-1} U' U'' + f_1 V'' - f_2 {U'}^2 - f_4 U'' + g_3 V' + g_5 U'+ g_6=0,
}
\end{array}
\right.
\label{eqsysUV}
\end{eqnarray} 
dont les onze coefficients ne doivent d\'ependre que de la variable r\'eduite $\xi$,
syst\`eme que nous allons d\'eterminer et int\'egrer en plusieurs \'etapes.

\'Etape 1.
Assignation de cinq des onze coefficients \`a des valeurs num\'eriques.
En effet, l'invariance de forme de (\ref{eqsysUV}) par la transformation affine
\`a cinq fonctions ajustables
\begin{eqnarray}
{\hskip -12.0truemm}
& & (U,V,\xi) \to (\UU,\VV,\xxi):\
U=\lambda_U(\xi) \UU(\xxi)+\mu_U(\xi),
V=\lambda_V(\xi) \VV(\xxi)+\mu_V(\xi), \xxi=f(\xi),
\label{eq-Affine}
\end{eqnarray}
permet, par une technique classique 
\cite[\S 14 p 223]{PaiBSMF}
d\'etaill\'ee par Bureau 
\cite[\S 20]{BureauMI},
d'assigner cinq des coefficients de (\ref{eqsysUV})
\`a des valeurs num\'eriques,
le meilleur choix \'etant celui qui simplifie le plus 
les calculs ult\'erieurs.
Un des choix permis est $\cu=1,\cv=1$ pour les coefficients de $U' U''$ et $U' V''$,
compl\'et\'e par l'annulation de trois autres coefficients,
\begin{eqnarray}
& & \cu=1,\cv=1,f_2=f_3=f_4=0.
\label{eqGauges}
\end{eqnarray}



\'Etape 2.
En supposant les six coefficients $f_j,g_j$ restants fonctions de $\xi$ seulement,
g\'en\'eration des conditions n\'ecessaires entre ces $(f_j,g_j)(\xi)$ 
pour que le syst\`eme  d'EDOs (\ref{eqsysUV}) poss\`ede la propri\'et\'e de Painlev\'e,
puis r\'esolution de ces conditions.
     			
\'Etape 3.
Preuve de la suffisance des conditions n\'ecessaires pr\'ecit\'ees,
par la d\'etermination de deux int\'egrales premi\`eres de (\ref{eqsysUV}),
abaissant ainsi l'ordre de quatre \`a deux,
suivie de l'int\'egration explicite de ce syst\`eme d'EDOs d'ordre deux.

\'Etape 4.
S\'election, parmi tous les syst\`emes (\ref{eqsysUV}) ainsi retenus et int\'egr\'es,
de ceux qui sont effectivement des r\'eductions de (\ref{eqsysuv}),
selon que l'int\'egration du syst\`eme d'EDPs aux trois inconnues $(\du,\dv,\xi)(x,y,t)$
est possible ou impossible. 

\subsection{\'Etape 2. Test de Painlev\'e du syst\`eme d'EDOs (\ref{eqsysUV})} 

Les deux familles de p\^oles simples mobiles du syst\`eme (\ref{eqsysUV}) 
\begin{eqnarray}
& & {\hskip -7.0truemm}
U' = \chi(\xi)^{-1} \left(i\nu + \sum_{j=1}^{+\infty} U_j(\xi) \chi^j\right),\ 
V' = \chi(\xi)^{-1} \left(1    + \sum_{j=1}^{+\infty} V_j(\xi) \chi^j\right),\ \chi'=1,\
\end{eqnarray} 
ont chacune les m\^emes indices de Fuchs \cite[\S 2.1.1]{CMBook2} $-1,0,0,1,2,4$ que l'EDP (\ref{eqsysuv}),
indices dont les trois premiers repr\'esentent les origines arbitraires de $\xi,U,V$.
Les trois suivants ne repr\'esentent les coefficients arbitraires
$V_1(\xi),V_2(\xi),V_4(\xi)$ 
que si la s\'erie ne contient pas de logarithmes mobiles.
Ces conditions n\'ecessaires s'\'ecrivent,
\begin{eqnarray}
& & {\hskip -15.0truemm}
Q_1\equiv P_1(f_j,g_j)=0,\
Q_2\equiv P_2(f_j,g_j)=0,\
Q_4\equiv (V_1'+V_2) Q_4^{(1)} + V_1 Q_4^{(2)} +  Q_4^{(3)}=0,\
\end{eqnarray}  
o\`u $P_1,P_2,Q_4^{(k)}$ d\'esignent des polyn\^omes diff\'erentiels de $f_j,g_j$.
Afin de laisser arbitraires $V_1$ et $V_2$,
dix conditions (cinq pour chaque signe de $i$) doivent donc \^etre v\'erifi\'ees.
Seules six d'entre elles sont non-nulles,
%
\begin{eqnarray}
& &
\left\lbrace
\begin{array}{ll}
\displaystyle{
Q_2 \equiv (g_3-f_1') \pm i g_5 =0,
}\\ \displaystyle{
%
%
%
%
Q_4^{(2)} \equiv (f_1'' + 2 f_1 f_1') \pm  i (f_5' + 2 f_1 f_5) =0,
}\\ \displaystyle{
Q_4^{(3)} \equiv \left(g_6' + 2 f_1 g_6 - \frac{1}{2} f_5^2  \right) 
           \pm i \left(f_6' + 2 f_1 f_6 + \frac{1}{2} f_5 f_1'\right)=0,
}
\end{array}
\right.
\label{eqQall}
\end{eqnarray} 
\'equivalentes \`a
\begin{eqnarray}
& & {\hskip -10.0truemm}
\left\lbrace
\begin{array}{ll}
\displaystyle{
g_3=f_1',
g_5=0,
f_1' + f_1^2 - k^2=0,
}\\ \displaystyle{
\left(\frac{\D}{\D \xi} + 2 f_1\right) f_5=0,
\left(\frac{\D}{\D \xi} + 2 f_1\right) g_6 - \frac{1}{2} f_5^2=0,
\left(\frac{\D}{\D \xi} + 2 f_1\right) f_6 + \frac{1}{2} f_5 f_1'=0,
}
\end{array}
\right.
\end{eqnarray} 
$k$ \'etant une constante d'int\'egration.

Ce syst\`eme consiste en une EDO non-lin\'eaire (pour $f_1$)
et une EDO lin\'eaire, lin\'earis\'ee de la pr\'ec\'edente,
avec trois seconds membres.
Sa solution g\'en\'erale d\'epend de quatre constantes arbitraires $k,\Kcinq,\Ksix,\Ksept$
(outre l'origine de $\xi$),
\begin{eqnarray}
& & {\hskip -13.0truemm}
g_3=f_1',
g_5=0,
f_5=(4 \Kcinq f_1)',
f_6=(4 \Ksix  f_1-   \Kcinq   f_1^2)',
g_6=(4 \Ksept f_1+ 4 \Kcinq^2 f_1^2)',
\label{eqnolog-sol}
\end{eqnarray} 
relations qui d\'efinissent le cas g\'en\'erique $k f_1' \not=0$,
\begin{eqnarray}
& & {\hskip -10.0 truemm} 
\left\lbrace
\begin{array}{ll}
\displaystyle{
%
f_1=k \coth(k \xi), 
g_3=k^2 (1-\coth^2 (k \xi)),
g_5=0,
f_5=k^2 (1-\coth^2 (k \xi))     (2 \Kcinq),
}\\ \displaystyle{
f_6=k^2 (1-\coth^2 (k \xi))\left[2 \Ksix -   \Kcinq   k \coth(k \xi)\right],
}\\ \displaystyle{
g_6=k^2 (1-\coth^2 (k \xi))\left[2 \Ksept+ 2 \Kcinq^2 k \coth(k \xi)\right],
}
\end{array}
\right.
\label{eqnolog-tri}
\end{eqnarray} 
et trois cas non-g\'en\'eriques $k f_1' \to 0$~:
\begin{eqnarray}
& & {\hskip -16.0 truemm} 
%
f_1= \frac{1}{\xi},
g_3=-\frac{1}{\xi^2},
g_5=0,
f_5=- 2 \frac{\Kcinq}{\xi^2}, 
f_6=- 2 \frac{\Ksix }{\xi^2}+ \frac{\Kcinq}  {\xi^3}\ccomma
g_6=- 2 \frac{\Ksept}{\xi^2}-2\frac{\Kcinq^2}{\xi^3}\ccomma
\label{eqnolog-rat}
\end{eqnarray} 
\begin{eqnarray}
& & {\hskip -16.0 truemm} 
f_1=k \not=0, 
g_3=
g_5=0,
f_5=- 8 \Kcinq k^2 e^{-2 k \xi},
f_6=- 8 \Ksix  k^2 e^{-2 k \xi},
g_6=- 8 \Ksept k^2 e^{-2 k \xi}- 16 \Kcinq^2 k^3 e^{-4 k \xi}.
\label{eqnolog-exp}
\end{eqnarray} 
et 
\begin{eqnarray}
& & {\hskip -16.0 truemm} 
f_1=0, 
g_3=
g_5=0,
f_5=- 8 \Kcinq,
f_6=- 8 \Ksix,
g_6=- 8 \Ksept + 32 \Kcinq^2 \xi.
\label{eqnolog-zer}
\end{eqnarray} 

Dans les solutions (\ref{eqnolog-tri}) et (\ref{eqnolog-rat}),
l'invariance de (\ref{eqsysUV}) 
par la translation $(V',\Ksix)$ $\to$ $(V'-2 \Ksept,\Ksix + 2 \Kcinq \Ksept)$
permet d'annuler $\Ksept$. 

\subsection{\'Etape 3.
Syst\`emes r\'eduits et leurs int\'egrales premi\`eres} 
\label{section-Systemes-reduits}

Une fois d\'etermin\'ees
les valeurs (\ref{eqnolog-tri})--(\ref{eqnolog-zer}) des coefficients $f_j, g_j$,
le syst\`eme (\ref{eqsysUV}) admet deux int\'egrales premi\`eres polynomiales
en $(U'',V'',U',V')$.
Puisque les indices de Fuchs non encore repr\'esent\'es par des arbitraires
sont $2$ et $4$, ces deux int\'egrales premi\`eres ont n\'ecessairement pour degr\'es de singularit\'e 
$2$ et $4$.
Leurs termes sont qualitativement fournis par les coefficients de $\lambda^{2}$
et de $\lambda^{4}$ du d\'eveloppement de Taylor de la fonction g\'en\'eratrice
\begin{eqnarray}
& &
\frac{1}{(1-\lambda^2 U'')(1-\lambda^2 V'')(1-\lambda   U')(1-\lambda   V')}
=\sum_{j=0}^{+\infty} \lambda^j P_j(U'',V'',U',V').
\end{eqnarray} 


Les syst\`emes r\'eduits et leurs int\'egrales premi\`eres sont ainsi les suivants.

Syst\`eme g\'en\'erique (\ref{eqnolog-tri})
(apr\`es une translation de $V'$ annulant $\Ksept$),
\begin{eqnarray}
& & {\hskip -8.0truemm}
\left\lbrace
\begin{array}{ll}
\displaystyle{	
%
%
%
%
%
U''' + 2 U' V'' + k \coth(k \xi) U''
}\\ \displaystyle{
\phantom{12345}
+ \nu^{-1} k^2 (1-\coth^2(k \xi)) (2 \Kcinq V' + 2 \Ksix - \Kcinq k \coth(k \xi))=0,
}\\ \displaystyle{
V''' -2 \nu^2 U' U'' + k \coth(k \xi) V'' + k^2 (1-\coth^2(k \xi)) \left(V'+ 2 \Kcinq^2 k \coth(k \xi)\right)=0,
}\\ \displaystyle{
K_2=- \nu^2 {U'}^2 + V''  + k \coth(k \xi) V' + \Kcinq^2 k^2 \coth^2(k \xi),
}\\ \displaystyle{			
K_4= \left(\nu \frac{\sinh(k \xi)}{k} U'' - \Kcinq \frac{k}{\sinh(k \xi)} \right)^2 
   + \left(\frac{\sinh(k \xi)}{k}{V''}\right)^2 
}\\ \displaystyle{			
\phantom{1234}
	- 4 \Kcinq \nu U' V' - {V'}^2 -4 \Ksix \nu U' 
  - 4 \Kcinq k \coth(k \xi) (\Kcinq V' + \Ksix),
}
\end{array}
\right.
\label{sysUVtri}
\end{eqnarray}
et sa limite rationnelle $f_1=1/\xi$ (\ref{eqnolog-rat})
(apr\`es une translation de $V'$ annulant $\Ksept$), 
\begin{eqnarray}
& & {\hskip -8.0truemm}
\left\lbrace
\begin{array}{ll}
\displaystyle{
%
%
U''' + 2 U' V'' + \frac{U''}{\xi} -2 \nu^{-1}(\Kcinq V' + \Ksix)\frac{1}{\xi^2} + \nu^{-1} \frac{\Kcinq}{\xi^3}=0,
}\\ \displaystyle{
V''' -2 \nu^2 U' U'' + \frac{V''}{\xi} - V' \frac{1}{\xi^2} - 2 \frac{\Kcinq^2}{\xi^3}=0\ccomma 
}\\ \displaystyle{
K_2=- \nu^2 {U'}^2 + V'' + V' \frac{1}{\xi} + \frac{\Kcinq^2}{\xi^2}\ccomma
}\\ \displaystyle{
K_4= \left(\nu \xi U'' - \frac{\Kcinq}{\xi} \right)^2 + (\xi V'')^2 - 4 \nu \Kcinq U' V' - {V'}^2 
  - 4 \nu \Ksix U' - 4 \Kcinq (\Kcinq V'+\Ksix)  \frac{1}{\xi}\cdot
}
\end{array}
\right.
\label{sysUVrat}
\end{eqnarray}
Les deux autres cas non-g\'en\'eriques sont
$f_1=k\not=0$ (\ref{eqnolog-exp}),
\begin{eqnarray}
& &
\left\lbrace
\begin{array}{ll}
\displaystyle{
%
%
U''' + 2 U' V'' + k U'' -8 \nu^{-1} e^{-2 k \xi} (\Kcinq k V' + \Ksix)=0,
}\\ \displaystyle{
V''' -2 \nu^2 U' U'' + k V'' - 16 \Kcinq^2 k e^{-4 k \xi} - 8 \Ksept k e^{-2 k \xi}=0,
}\\ \displaystyle{
K_2=- \nu^2 {U'}^2 + V'' + k V' + 4 \Kcinq^2 e^{-4 k \xi} + 4 \Ksept e^{-2 k \xi}, 
}\\ \displaystyle{
K_4= e^{2 k \xi} \left(\nu^2 {U''}^2 + {V''}^2\right) -8 \nu \Kcinq k (U'' + 2  U' V') 
 - 8 \nu (2 \Ksix + \Kcinq k^2) U'
}\\ \displaystyle{
\phantom{1234}
 - 16 \Ksept k V' -32 \Kcinq (\Kcinq k V' + \Ksix) e^{-2 k \xi},
}
\end{array}
\right.
\label{sysUVexp}
\end{eqnarray}
et $f_1=0$ (\ref{eqnolog-zer}), 
\begin{eqnarray}
& &
\left\lbrace
\begin{array}{ll}
\displaystyle{
%

%
U''' + 2 U' V'' -8 \nu^{-1} (\Kcinq V' + \Ksix)=0,
}\\ \displaystyle{
V''' -2 \nu^2 U' U'' + 32 \Kcinq^2 \xi - 8 \Ksept=0,
}\\ \displaystyle{
K_2=- \nu^2 {U'}^2 + V'' + 16 \Kcinq^2 \xi^2 -8 \Ksept \xi, 
}\\ \displaystyle{
K_4= \nu^2 {U''}^2 + {V''}^2 -8 \nu \Kcinq (U'' + 2  U' V') 
 - 16 \nu \Ksix U'
}\\ \displaystyle{
\phantom{1234}
 (64 \Kcinq^2 \xi -16 \Ksept) V' + 64 \Kcinq \Ksix \xi.
}
\end{array}
\right.
\label{sysUVzer}
\end{eqnarray}

\textit{Remarque}.
Il n'existe pas de changement de la variable ind\'ependante 
rendant autonomes les deux premi\`eres \'equations de (\ref{sysUVtri}).
Les seuls syst\`emes autonomes sont
(\ref{sysUVrat}) pour $\Ksix=0$ (apr\`es le changement $\xi \to \log(\xi)$) et  
(\ref{sysUVexp}),
(\ref{sysUVzer}) pour certaines valeurs des param\`etres.

Certains de ces syst\`emes autonomes ont d\'ej\`a \'et\'e mentionn\'es par Lou Sen-yue, ce sont~:
(\ref{sysUVrat}) avec la  contrainte  suppl\'ementaire  $\Kcinq=0$        \cite[Eqs.~(79)--(80)]{Lou2000}, 
(\ref{sysUVexp}) avec les contraintes suppl\'ementaires $\Kcinq=\Ksept=0$ \cite[Eqs.~(63)--(64)]{Lou2000}, 
et bien s\^ur la d\'eg\'en\'erescence homog\`ene $U''' + 2 U' V'' =0, V''' -2 \nu^2 U' U''=0$ \cite[Eqs.~(55)--(57)]{Lou2000} 
commune aux quatre syst\`emes.
Voir \'egalement \cite{CMM1996},
qui consid\`ere un syst\`eme l\'eg\`erement diff\'erent.

\subsection{\'Etape 3, suite.
Int\'egration des syst\`emes r\'eduits} 
\label{section-Integration}



L'\'elimination de $V'$ entre $K_2$ et $K_4$ engendre
une EDO d'ordre deux en $U'$,
dont le degr\'e en $U'''$ est 
deux pour (\ref{sysUVtri}) et (\ref{sysUVexp}), 
un   pour (\ref{sysUVrat}) et (\ref{sysUVzer}),
et qui, en tant que r\'eduction non-caract\'eristique d'une EDP ayant la propri\'et\'e de Painlev\'e,
poss\`ede \'egalement cette propri\'et\'e.
Or, toutes ces EDOs (ordre deux et degr\'e un ou deux)
ont d\'ej\`a \'et\'e \'enum\'er\'ees et int\'egr\'ees  
par 
    Gambier \cite{GambierThese},
		Chazy    \cite{ChazyThese},
    Bureau   \cite{BureauMIII} et
    Cosgrove \cite{CosScou,Cos2006b},
il suffit donc de rechercher dans leurs tables.
\medskip

Pour le degr\'e deux, qui contient le cas g\'en\'erique,
et dans les cas non-autonomes, 
il suffit d'\'etablir la d\'ecomposition de Gauss de la forme quadratique de $U'''$,
\begin{eqnarray}
& & {\hskip -8.0truemm}
P_1^2(U''',U'',U';\xi) + P_0^2(U';\xi) P_2(U'',U';\xi)=0,
\nonumber \\ & & {\hskip -8.0truemm}
P_1 \equiv U''' + a_1(\xi) U'' + c_3(\xi) {U'}^3 + c_2(\xi) {U'}^2 + c_1(\xi) U' + c_0(\xi),
P_2 \equiv {U''}^2 + \dots,
\end{eqnarray} 
les $P_j$ d\'esignant des polyn\^omes de degr\'e $j$ en $U'''$,
\`a coefficients fonctions de $U'',U'$, $\xi$.
Les seuls candidats non-autonomes dans les tables classiques sont alors (voir l'appendice \ref{sectionCn})~: 
$\CVI$ et $\CVb$ pour $c_3 \not=0$,  
$\CVa$ et $\CIV$ pour $c_3     =0,c_2\not=0$, 
$\CIII$          pour $c_3=c_2=0$,
la nature du carr\'e $P_0^2$
r\'eduisant le choix \`a 
$\CVI$ pour $c_3 \not=0$,  
$\CVa$ pour $c_3     =0,c_2\not=0$.

\medskip

Pour le degr\'e un,   
on obtient ainsi 
$\PV$   (cas $K_2\not=0$ de (\ref{sysUVrat})),
$\PIII$ (cas $K_2=0$     de (\ref{sysUVrat})),
$\PIV$  (cas $K_5\not=0$ de (\ref{sysUVzer})) et
$\PII$  (cas $K_5=0$     de (\ref{sysUVzer})), 
donc toutes les $\Pn$ 
(rappelons que $\PI$ et $\PII$ peuvent \^etre rassembl\'ees en une seule \'equation 
\cite{PaiCRAS1898-126-1697})
sauf $\PVI$.
\medskip

La liste compl\`ete des solutions $U(\xi)$ en fonction de 
$\Cn(u,x,d_1,d_2,d_3,d_4)$ ou 
\hfill\break\noindent
$\Pn(u,x,\alpha,\beta,\gamma,\delta)$
est la suivante.

\begin{enumerate}
\item
Syst\`eme (\ref{sysUVtri}). Transform\'ee affine de $\CVI$,
\begin{eqnarray}
& & {\hskip -13.0 truemm}
%
\left\lbrace
\begin{array}{ll}
\displaystyle{
\nu \frac{\D U}{\D \xi}= k u - \Kcinq k \coth(k \xi),\ \frac{\D x}{\D \xi}=\frac{k}{\sinh(k \xi)},
}\\ \displaystyle{
\fVI(x)= i \cosh(k \xi)=i \coth(x), \gVI(x)=\coth(k \xi)=\cosh(x),
}\\ \displaystyle{
\Kcinq=k d_1,
K_2=k^4 \left(\frac{d_2}{2}+2 d_1^2\right),
\Ksix=k^4 \frac{d_3}{2},
K_4=k^6 \left(d_4 -\left(\frac{d_2}{2}+2 d_1^2\right)^2\right).
}
\end{array}
\right.
\label{eqsysUVtrisol}
\end{eqnarray}

\item
Syst\`eme (\ref{sysUVrat}), $K_2\not=0$. Transform\'ee homographique de $\PV$,
\begin{eqnarray}
& & {\hskip -13.0 truemm}
\left\lbrace
\begin{array}{ll}
\displaystyle{
%
\nu \frac{\D U}{\D \xi}=-\frac{\Kcinq}{\xi} + \rKtwo\frac{1+u}{1-u}, x=\xi,
}\\ \displaystyle{
  K_2=-\rKtwo^2 = \frac{\delta}{2}\not=0,
4 K_4=8(\alpha-\beta)\delta+ \gamma^2,
  \Ksix=-(\alpha+\beta) \rKtwo,
	\Kcinq=\frac{\rKtwo \gamma}{2 \delta}.
}
\end{array}
\right.
\end{eqnarray}

\item
Syst\`eme (\ref{sysUVrat}), $K_2=0$. Transform\'ee affine de $\PIII$,
\begin{eqnarray}
& & {\hskip -15.0truemm}
%
\nu \frac{\D U}{\D \xi}= \frac{\ka u - \Kcinq}{\xi}, x= \xi,
\ka^2=-\frac{\gamma}{4}\not=0,
 8 \Kcinq=-\frac{\alpha}{\ka},
16 K_4=\gamma \delta,
 8 \Ksix=-\beta \ka.
\end{eqnarray}

\item
Syst\`eme (\ref{sysUVexp}), $\Kcinq\not=0$. Transform\'ee affine de $\CVa$,
\begin{eqnarray}
& & {\hskip -17.0 truemm}
\left\lbrace
\begin{array}{ll}
\displaystyle{
%
\nu \frac{\D U}{\D \xi}= -2 \frac{k_0^2}{\Kcinq} u - 2 \Kcinq e^{-2 k \xi}, x=-2 \frac{k_0}{k} e^{-k \xi},
}\\ \displaystyle{
k_0^2=-i k \Kcinq,
\Ksept=-\frac{i}{4} d_2 k \Kcinq,
\Ksix=-((d_3+1) k^2 +2 K_2) \frac{\Kcinq}{2},
}\\ \displaystyle{
K_4=-4 i (d_4 k^2 + K_2 d_2) k \Kcinq.
}
\end{array}
\right.
\label{eqReduitCVa}
\end{eqnarray}

\item
Syst\`eme (\ref{sysUVexp}), $\Kcinq=0$. Transform\'ee affine de $\CIII$,
\begin{eqnarray}
& & {\hskip -21.0 truemm}
%
\nu \frac{\D U}{\D \xi}= i k u, x=-2 \frac{k_0}{k} e^{-k \xi},
\Ksept= \frac{k_0^2}{4} d_2,
\Ksix= i k \frac{k_0^2}{2} d_3,
K_4=-4 k_0^2 (d_4 k^2 + K_2 d_2).
\end{eqnarray}

\item
Syst\`eme (\ref{sysUVzer}), $\Kcinq\not=0$. Transform\'ee affine de $\PIV$,
\begin{eqnarray}
& & {\hskip -18.0 truemm} 
\left\lbrace
\begin{array}{ll}
    \displaystyle{ 
\nu \frac{\D U}{\D \xi}=- \frac{\Ksept}{\Kcinq} + i \mu u,
\xi= \frac{x}{\mu} + \frac{\Ksept}{4 \Kcinq^2},
\mu^2=-4 i \Kcinq,
}\\ \displaystyle{ 
\alpha=i \frac{K_2 \Kcinq^2 + \Ksept^2}{8 \Kcinq^3}\ccomma
\beta=\frac{(K_2^2-K_4) \Kcinq^4 +16 \Ksix \Ksept \Kcinq^3 + 2 \Ksept^2 \Kcinq^2 + \Ksept^4}{32 \Kcinq^6}-\frac{1}{2}  \cdot
}
\end{array}
\right.
\end{eqnarray} 

\item
Syst\`eme (\ref{sysUVzer}), $\Kcinq=0, \Ksept\not=0$. Transform\'ee affine de $\PII$,
\begin{eqnarray}
& & {\hskip -10.0 truemm} 
\nu \frac{\D U}{\D \xi}=\lambda u,
\xi= \frac{x}{\mu} - \frac{K_2}{8 \Ksept},
\lambda=i \nu^{-1} \mu,
\mu^3=- 16 \Ksept,
\alpha=\frac{i \Ksix}{2 \Ksept}\cdot
\end{eqnarray} 

\item
Syst\`eme (\ref{sysUVzer}), $\Kcinq=0, \Ksept=0$. Fonction elliptique d'ordre deux,
\begin{eqnarray}
& &  {\hskip -10.0 truemm} 
\nu^2 {U''}^2 + \nu^4 {U'}^4 + 2 \nu^2 K_2 {U'}^2 -16 \nu \Ksix  U' + K_2 - K_4=0.
\end{eqnarray}

\item
Tous syst\`emes, cas o\`u les deux derni\`eres \'equations ($K_2=\dots$, $K_4=\dots$)
sont autonomes. Fonction elliptique d'ordre deux.

\end{enumerate}

\textit{Remarque}.
Dans tous les cas \'enum\'er\'es ci-dessus, tous les param\`etres des $\Cn$ ou des $\Pn$
sont arbitraires.
L'exigence que (\ref{eqsysUV}) soit une r\'eduction de (\ref{eqsysuv}) peut cr\'eer des
contraintes entre ces param\`etres, comme d\'ecrit dans la section suivante.

\section{\'Etape 4.
R\'eductions non-caract\'eristiques admises} 
\label{sectionReductionsPDE}


Il reste \`a d\'eterminer,
pour chacun des quatre syst\`emes r\'eduits 
(\ref{sysUVtri}), (\ref{sysUVrat}), (\ref{sysUVexp}), (\ref{sysUVzer}), 
la variable $\xi(x,y,t)$ de la r\'eduction d\'efinie par (\ref{eqred-no-UV}) avec $\cu=\cv=1$,
les deux coefficients $(\du,\dv)(x,y,t)$,
ainsi que les contraintes entre les constantes d'int\'egration $\Kcinq$, $\Ksix$, $\Ksept$
du syst\`eme (\ref{eqnolog-sol}).
\medskip

Les trois fonctions $(\xif,\du,\dv)(x,y,t)$ ob\'eissent \`a un syst\`eme de sept EDPs,
\begin{eqnarray}
& & {\hskip -10.0 truemm} 
\left\lbrace
\begin{array}{ll}
         \displaystyle{ (\xif_x \xif_y)_t=0,                                             
}\\      \displaystyle{ \xif_{xy} -  \xif_x \xif_y     f_1(\xif)=0,                      
}\\ \displaystyle{ \du_{,x} \xif_y    + \du_{,y} \xif_x=0,                               
}\\ \displaystyle{ \du_{,x} \xif_{yt} + \du_{,y} \xif_{xt} - \xif_x \xif_y \xif_t f_5(\xif)=0,  
}\\ \displaystyle{ \dv_{,x} \xif_y + \dv_{,y} \xif_x + \xif_x \xif_y \xif_t f_1'(\xif)=0, 
}\\ \displaystyle{ \du_{,xyt} + \du_{,x} \dv_{,y} + \du_{,y} \dv_{,x}- \xif_x \xif_y \xif_t f_6(\xif)=0, 
}\\ \displaystyle{ \dv_{,xy} - (\du_{,x} \du_{,y})_t - \xif_x \xif_y \xif_t g_6(\xif)=0,  
}
\end{array}
\right.
\label{eqsyszdudv}
\label{eqsysxidudv9}
\end{eqnarray}
contraint par $\xif_x \xif_y \xif_t \not=0$,                                  
invariant par les transformations conformes
$(x,y,t) \to (\varphi_1(x)$, $\varphi_2(y)$, $\varphi_3(t))$,
et dont les coefficients $f_j,g_j$ sont d\'efinis par (\ref{eqnolog-sol}).

Seules les deux premi\`eres \'equations, 
non-lin\'eaires en $\xif(x,y,t)$,
pr\'esentent quelque difficult\'e, elles sont r\'esolues ci-apr\`es. 
Les quatre suivantes d\'efinissent
les coefficients $\du$ et $\dv$ par leur gradient
\begin{eqnarray}
& & {\hskip -10.0 truemm} 
\left\lbrace
\begin{array}{ll}
    \displaystyle{ 
\du_{,x} =  \frac{\xif_x^2 \xif_y \xif_t f_5(\xif)}{\xif_x \xif_{yt} - \xif_y \xif_{xt}}\ccomma
}\\ \displaystyle{ 
\du_{,y} =  \frac{\xif_y^2 \xif_x \xif_t f_5(\xif)}{\xif_y \xif_{xt} - \xif_x \xif_{yt}}\ccomma
}\\ \displaystyle{
     \dv_{,x}=\frac{\xif_x \xif_{yt} - \xif_y \xif_{xt}}{2 \xif_y f_5(\xif)}
      \left[\frac{\du_{,xyt}}{\xif_x \xif_y \xif_t} 
			      - f_6(\xif)\right]
			-\frac{\xif_x \xif_t}{2} f_1'(\xif),
}\\ \displaystyle{
     \dv_{,y}=\frac{\xif_y \xif_{xt} - \xif_x \xif_{yt}}{2 \xif_x f_5(\xif)}
      \left[\frac{\du_{,xyt}}{\xif_x \xif_y \xif_t }
			      - f_6(\xif)\right]
			-\frac{\xif_y \xif_t}{2} f_1'(\xif),
}
\end{array}
\right.
\label{eq-f5nonzero}
\end{eqnarray} 
et la derni\`ere \'equation (\ref{eqsyszdudv})${}_7$
cr\'ee des contraintes entre les diverses constantes d'int\'egration.

Quant aux \'eventuelles solutions $\xi$ annulant le wronskien,
\begin{eqnarray}
& &  {\hskip -10.0 truemm} 
\xif_x \xif_{yt} - \xif_y \xif_{xt}=0, \xif_x  \xif_y  \xif_t \not=0, 
\label{eqWronskien-nul}
\end{eqnarray}
elles seront consid\'er\'ees s\'epar\'ement.

Revenons aux deux \'equations (\ref{eqsyszdudv})${}_1$, (\ref{eqsyszdudv})${}_2$.
Pour chaque valeur (\ref{eqnolog-tri})--(\ref{eqnolog-zer}) de $f_1$,
il existe deux fonctions d'une variable, $\xiofZ(Z)$ et $\psiofZ(Z)$,
d\'efinies par les EDOs,
\begin{eqnarray}
& &  {\hskip -10.0 truemm} 
\xiofZ''(Z) - f_1(\xiofZ(Z)) {\xiofZ'}^2(Z)=0,
\psiofZ''' - 2 \frac{\xiofZ''}{\xiofZ'} \psiofZ''=0,
\end{eqnarray}
qui transforment le couple (\ref{eqsyszdudv})${}_1$--(\ref{eqsyszdudv})${}_2$
en un couple \'equivalent du type de d'Alembert,
\begin{eqnarray}
& &  {\hskip -10.0 truemm} 
Z_{xy}=0, \left[\psiofZ(Z)\right]_{xyt}=0. 
\label{eqdAlembert}  
\end{eqnarray}
Leurs valeurs sont donn\'ees par le tableau 
\begin{eqnarray}
& & {\hskip -10.0 truemm} 
\begin{array}{l|l|l|l}
& f_1                      & \xi=\xiofZ(Z)                      & \psiofZ(Z)                   \cr
%
&k\coth(k\xi)              &\displaystyle\frac{1}{k}\log\coth(Z)&\displaystyle\log \sinh(2 Z)  \cr 
&\displaystyle\frac{1}{\xi}&e^{\displaystyle Z}                 &e^{\displaystyle        2 Z}  \cr
&k \not=0                  &\displaystyle-\frac{1}{k}\log(Z)    &\log (Z)                      \cr
&0                         &Z                                   &Z^2.                          \cr
\end{array}
\label{eqphipsi}
\end{eqnarray}           
\begin{itemize}
	\item 
Pour $f_1=k \coth(k \xi)$ (resp.~$f_1=k\not=0$), 
il existe une troisi\`eme fonction $\FG$ d'une variable\footnote{%
Due \`a Wolfgang Schief,
cette repr\'esentation permet d'extrapoler \`a deux fonctions suppl\'ementaires de $t$
une solution particuli\`ere que nous avions obtenue.
},
qu'il est loisible de d\'efinir par
\begin{eqnarray}
& &  {\hskip -10.0 truemm} 
Z(x,y,t)=\frac{1}{4}\log(\FG(\fa(x,t)))+\frac{1}{4}\log(\FG(\fb(y,t))),
\label{eqdefPhi}  
\end{eqnarray}
et qui transforme $\left[\psiofZ(Z)\right]_{xyt}=0$ en
\begin{eqnarray}
& &  {\hskip -10.0 truemm} 
\lbrack\log(\fa(x,t)+\fb(y,t))\rbrack_{xyt}=0,
\label{pdeab}
\end{eqnarray}
\'equation fonctionnelle 
dont la solution g\'en\'erale d\'epend de cinq fonctions arbitraires
d'une variable (Appendice \ref{section-xi}).
\smallskip
Solution des syst\`emes respectifs
\begin{eqnarray}                
& & {\hskip -10.0 truemm} 
\left\lbrace
\begin{array}{ll}
    \displaystyle{ 
(a+b)(1-\FG(a) \FG(b)) \FG''(a)+2(a+b)\FG(b){\FG'(a)}^2+2(1-\FG(a) \FG(b))\FG'(a)=0,
}\\ \displaystyle{
(a+b)(1-\FG(a) \FG(b)) \FG''(b)+2(a+b)\FG(a){\FG'(b)}^2+2(1-\FG(a) \FG(b))\FG'(b)=0,
}
\end{array}
\right.
\end{eqnarray}
et
\begin{eqnarray}                
& & {\hskip -10.0 truemm} 
\left\lbrace
\begin{array}{ll}
    \displaystyle{ 
-(a+b)(\FG(a)+\FG(b)) \FG''(a)+2(a+b){\FG'(a)}^2-2(\FG(a)+\FG(b))\FG'(a)=0,
}\\ \displaystyle{
-(a+b)(\FG(a)+\FG(b)) \FG''(b)+2(a+b){\FG'(b)}^2+2(\FG(a)+\FG(b))\FG'(b)=0,
}
\end{array}
\right.
\end{eqnarray} 
cette fonction $\FG$ est une homographie car son schwarzien est nul,
\begin{eqnarray}
& &  {\hskip -10.0 truemm} 
\forall X: \left\lbrace \FG;X\right\rbrace=0,
\end{eqnarray}
conduisant aux expressions finales respectives,
\begin{eqnarray}
& &  {\hskip -10.0 truemm} 
f_1=k\coth(k\xi):\
\xi=\frac{1}{k}\log\coth(Z),
Z=\frac{1}{4} \log \frac{1+\fa(x,t)}{1-\fa(x,t)} + \frac{1}{4} \log \frac{1+\fb(y,t)}{1-\fb(y,t)}\ccomma
\end{eqnarray}
et
\begin{eqnarray}
& &  {\hskip -10.0 truemm} 
f_1=k:\
\xi=-\frac{1}{k}\log(Z),
Z=\fa(x,t)+\fb(y,t),
\end{eqnarray}
avec pour $\fa, \fb$ les valeurs (\ref{eqsolpdeab})--(\ref{eqsolpdeab-contraintes}).
	\item 
	Pour $f_1=1/\xi$ (resp.~$f_1=0$),
l'EDP $\left[\psiofZ(Z)\right]_{xyt}=0$ est \`a variables s\'epar\'ees,
\begin{eqnarray}                
& & {\hskip -14.0 truemm} 
f_1=1/\xi,
\xi=e^{\displaystyle Z},
Z=\fa(x,t)+\fb(y,t), 
\left[e^{2 Z}\right]_{xyt} \equiv e^{2 Z} \frac{([\log \fa_x + 2 \fa]+[\log \fb_y + 2 \fb])_t}{\fa_x \fb_y},
\end{eqnarray}
\begin{eqnarray}                
& & {\hskip -10.0 truemm} 
f_1=0,
\xi=Z,
Z=\fa(x,t)+\fb(y,t), 
\left[Z^{2}\right]_{xyt} \equiv 2 \fa_x \fb_y ([\log \fa_x]+[\log \fb_y])_t,
\end{eqnarray}
conduisant ainsi \`a une solution g\'en\'erale $Z$ qui d\'epend aussi de cinq fonctions arbitraires
d'une variable,
\begin{eqnarray}                
& & {\hskip -10.0 truemm} 
f_1=1/\xi:\
\left\lbrace
\begin{array}{ll}
    \displaystyle{ 
\log \fa_x + 2 \fa - \log h_1(t) - \log \farbx'(x)=0,
}\\ \displaystyle{
\log \fb_y + 2 \fb + \log h_2(t) - \log \farby'(y)=0,
}\\ \displaystyle{
Z=\frac{1}{2} \log    [(\fx(x)+h_1(t)) \farbt(t)]
 +\frac{1}{2} \log\frac{\fy(y)+h_2(t)}{\farbt(t)}\ccomma
}
\end{array}
\right.
\end{eqnarray}
\begin{eqnarray}                
& & {\hskip -10.0 truemm} 
f_1=0:\
\left\lbrace
\begin{array}{ll}
    \displaystyle{ 
\log \fa_x - \log \farbt(t) - \log \farbx'(x)=0,
}\\ \displaystyle{
\log \fb_y + \log \farbt(t) - \log \farby'(y)=0,
}\\ \displaystyle{
Z=     (\fx(x)+h_1(t)) \farbt(t)
 +\frac{\fy(y)+h_2(t)}{\farbt(t)}\cdot
}
\end{array}
\right.
\end{eqnarray}
\end{itemize}

\vfill\eject

Finalement,
les r\'eductions non-caract\'eristiques ainsi obtenues sont les suivantes.
Elles d\'ependent toutes de diverses fonctions arbitraires d'une variable~:
$\fx(x),\fy(y)$ (choisies \'egales \`a $x,y$ par l'invariance conforme),
$\eff_1(t),\eff_2(t),\eff_3(t)$,
ou encore (autre notation) 
$h_0(t),h_1(t),h_2(t)$.

\medskip

\textbf{Syst\`eme r\'eduit (\ref{sysUVtri}) ($f_1=k \coth(k \xi)$)}. 

Il n'existe pas de r\'eduction de wronskien nul.
\medskip
Les variables r\'eduites $(\xif,\du,\dv)$ les plus g\'en\'erales
\begin{eqnarray}
& & {\hskip -15.0truemm}
\left\lbrace
\begin{array}{ll}
\displaystyle{
\xi=\frac{1}{k}\log\frac{\sqrt{N_1 N_2}+\sqrt{D_1 D_2}}{\sqrt{N_1 N_2}-\sqrt{D_1 D_2}},
}\\ \displaystyle{
%
N_1=(1+\eff_1)(x-\eff_3) + \eff_2,
N_2=(1-\eff_1)(y-\eff_3) - \eff_2,
}\\ \displaystyle{
D_1=(1-\eff_1)(x-\eff_3) - \eff_2,
D_2=(1+\eff_1)(y-\eff_3) + \eff_2,
}\\ \displaystyle{
\du=
 \Kcinq \log\left\lbrack
 (\hsum+\hdif) \sqrt{\frac{D_2 N_1}{D_1 N_2}}
+(\hsum-\hdif) \sqrt{\frac{D_1 N_2}{D_2 N_1}}
\right\rbrack,
}\\ \displaystyle{
%
%
\dv=
         \frac{\hsum-\hdif}{4 (\eff_1+1)} \left(\frac{1}{N_1}+\frac{1}{D_2}\right)
        -\frac{\hsum+\hdif}{4 (\eff_1-1)} \left(\frac{1}{N_2}+\frac{1}{D_1}\right)
}\\ \displaystyle{
\phantom{\dv=}
				-2 (1-\eff_1^2) \cone \frac{(1-\eff_1) N_1+(1+\eff_1) N_2}{(\hsum+\hdif) D_2 N_1+(\hsum-\hdif) D_1 N_2}
}\\ \displaystyle{
\phantom{\dv=}
 - \frac{\Ksix}{2 k \Kcinq \eff_2}\frac{(\hsum+\hdif) D_2 N_1-(\hsum-\hdif) D_1 N_2}{\sqrt{N_1 D_1 N_2 D_2}}\ccomma
}\\ \displaystyle{
\hsum=\eff_2 \eff_1' -\eff_1 \eff_2'+(\eff_1^2+1) \eff_3',
\hdif=\eff_2'-2 \eff_1 \eff_3',
}\\ \displaystyle{
\cone=\frac{\hbox{polyn\^ome diff\'erentiel de } \eff_1,\eff_2,\eff_3 \hbox{ de 30 termes}}
           {(1+\eff_1^2)\eff_2\eff_1'+(1-\eff_1^2)\eff_1\eff_2'+(1-\eff_1^2)^2\eff_3'}\ccomma
}
\end{array}
\right.
\end{eqnarray}
sont assorties de deux contraintes,
\begin{eqnarray}
& & \Kcinq^2=-1, \Ksix=2 \Kcinq \Ksept, 
\end{eqnarray}
qui laissent arbitraires les quatre param\`etres de $\CVI$
d\'efinis en (\ref{eqsysUVtrisol}).
Le syst\`eme (\ref{eqsysuv}) admet donc, comme esp\'er\'e,
une r\'eduction \`a une transform\'ee alg\'ebrique de la $\PVI$
la plus g\'en\'erale.
\medskip
Un exemple simple d'une telle r\'eduction est fourni par
le choix $(\eff_1,\eff_2,\eff_3)=(0,t,t)$,
\begin{eqnarray}
& & {\hskip -15.0truemm}
\left\lbrace
\begin{array}{ll}
\displaystyle{
\xi=\frac{1}{k}\log\frac{\sqrt{x (y-2 t)}+\sqrt{y (x-2 t)}}
                        {\sqrt{x (y-2 t)}-\sqrt{y (x-2 t)}}\ccomma
f_1=k \frac{x y -t (x+y)}{\sqrt{x y (x-2 t)(y-2 t)}}\ccomma
}\\ \displaystyle{
\du=\frac{\Kcinq}{2} \log \frac{x y}{(x-2 t)(y-2 t)}\ccomma
}\\ \displaystyle{
\dv=-\frac{\Ksix \sqrt{x y}}{k \Kcinq t \sqrt{(x-2 t)(y-2 t)}}
 -\frac{4 t -x-y}{2 (x-2 t) (y+2 t)}\cdot
}
\end{array}
\right.
\end{eqnarray} 

\medskip

\textbf{Syst\`eme r\'eduit (\ref{sysUVrat}) ($f_1=1/\xi$)}.

Il n'existe pas de r\'eduction de wronskien nul.

Les variables de r\'eduction 
\begin{eqnarray}
& & {\hskip -15.0truemm}
\left\lbrace
\begin{array}{ll}
\displaystyle{
\xi=\sqrt{(x+h_1(t))(y+h_2(t))}, (h_1', h_2') \not=(0,0),
}\\ \displaystyle{
\du=\Kcinq \log[(x+h_1)h_2' -(y+h_2) h_1']- \frac{\Kcinq}{2} \log[(x+h_1)(y+h_2)], 
}\\ \displaystyle{
\dv=       \log[(x+h_1)h_2' -(y+h_2) h_1']- \frac{1}{4}      \log[(x+h_1)(y+h_2)] 
}\\ \displaystyle{
\phantom{12345}
		+\frac{K_6}{K_5} \sqrt{(x+h_1(t))(y+h_2(t))},
}
\end{array}
\right.
\label{eqsysratreduc}
\end{eqnarray}
requi\`erent les deux contraintes,
\begin{eqnarray}
& & (\Kcinq^2+1) (h_2' h_1'' - h_1' h_2'')=0, \Ksix=2 K_5 K_7, 
\end{eqnarray}
qui ne restreignent aucun des param\`etres de la $\PV$.

\textbf{Syst\`eme r\'eduit (\ref{sysUVexp}) ($f_1=k \not=0$)}. 

Le wronskien ne s'annule que pour $\eff_3'=0$.

Si $\eff_3'\not=0$ et $\Kcinq\not=0$, 
les six premi\`eres \'equations (\ref{eqsyszdudv})
d\'efinissent les variables de r\'eduction
\begin{eqnarray}
& & {\hskip -15.0truemm}
\left\lbrace
\begin{array}{ll}
\displaystyle{
\xi= \frac{1}{k}\log\frac{(x-\eff_3(t))(y-\eff_3(t))}{(x-y) \eff_2(t)}\ccomma\ \eff_3'\not=0,
}\\ \displaystyle{
\du=\Kcinq \eff_2 \left[
 2 \eff_2 \left(\frac{1}{x-\eff_3(t)}+\frac{1}{y-\eff_3(t)} \right)^2
 + 4 \frac{\eff_2}{\eff_3'} \left(\frac{1}{x-\eff_3(t)}+\frac{1}{y-\eff_3(t)} \right)
\right],
}\\ \displaystyle{
\dv=\frac{\Ksix+k \Kcinq}{k \Kcinq} \eff_3'  \left(\frac{1}{x-\eff_3(t)}+\frac{1}{y-\eff_3(t)} \right),
}
\end{array}
\right.
\label{eq-exp-xidudv}
\end{eqnarray}
mais la septi\`eme \'equation (\ref{eqsyszdudv})${}_7$ n'a alors aucune solution.
\medskip

Si $\eff_3'\not=0$ et $\Kcinq=0$, on obtient
\begin{eqnarray}
& & {\hskip -15.0truemm}
\left\lbrace
\begin{array}{ll}
\displaystyle{
\xi= \frac{1}{k}\log\frac{(x-\eff_3(t))(y-\eff_3(t))}{(x-y) \eff_2(t)}\ccomma\ \eff_3'\not=0,
}\\ \displaystyle{
\du= 0,
}\\ \displaystyle{
\dv=-\frac{4}{3 k} \Ksept \eff_2^2 \eff_3' p^3 - \frac{4}{k} \Ksept \eff_2' p^2 + \farbone(t) p,
p=\frac{1}{x-\eff_3(t)}+\frac{1}{y-\eff_3(t)}.
}
\end{array}
\right.
\end{eqnarray}
\medskip

Si $\eff_3'=0$,
les variables sont d\'efinies par
\begin{eqnarray}
& & {\hskip -15.0truemm}
\xi= \frac{1}{k}\log\frac{x y}{(x-y) \eff_2(t)}\ccomma\ \eff_2' \not=0,
\du=\Fu(p,t),
\dv=\Fvt(p,t),
p=\frac{x+y}{x y}\ccomma
\label{eq-exp-xidudv-Wronskien0}
\end{eqnarray}
\`a la condition que $\Kcinq$ soit nul et que
les fonctions $\Fu,\Fvt$ ob\'eissent au syst\`eme,
\begin{eqnarray}
& & {\hskip -15.0truemm}
\Fu_{ppt} + 2 \Fu_p \Fvt_p     + 8 \frac{\Ksix} {k} \eff_2 \eff_2'=0,
- 2 \Fu_p \Fu_{pt} + \Fvt_{pp} + 8 \frac{\Ksept}{k} \eff_2 \eff_2'=0.
\label{eqsysexpFuFvt}
\end{eqnarray}
\medskip

On en conclut d'une part qu'il n'existe pas de variables $(\xi,\du,\dv)$
engendrant la transform\'ee affine (\ref{eqReduitCVa}) de $\CVa$,
d'autre part qu'il existe bien une r\'eduction \`a la $\CIII$ la plus g\'en\'erale.
\medskip

En tant que r\'eduction non-caract\'eristique de (\ref{eqsysuv}), 
le syst\`eme (\ref{eqsysexpFuFvt}) poss\`ede la propri\'et\'e de Painlev\'e quels que soient
$\Ksix$ et $\Ksept$.
Bien que nous n'ayons pas r\'eussi \`a l'int\'egrer,
il est facile d'en trouver des solutions particuli\`eres.

Une premi\`ere telle solution est une solution elliptique,
solution g\'en\'erale de la r\'eduction
$\Fu(p,t)=\Fu_r(p-\eff_2^2)$.
Une deuxi\`eme solution particuli\`ere est sugg\'er\'ee par l'existence de 
deux p\^oles mobiles simples 
de r\'esidus oppos\'es pour le champ $\Fu_p$;
la troncature \`a une famille \cite{WTC}
\cite[\S 5.6.1]{CMBook2} de l'EDP pour $\Fu$
engendre une solution repr\'esent\'ee par
\begin{eqnarray}
& & {\hskip -15.0truemm}
\left\lbrace
\begin{array}{ll}
\displaystyle{
\Fu=i\log(\phiFu) - \frac{i}{2} \log(\phiFu_p),
}\\ \displaystyle{
(k \Ksept + i \Ksix)\left\lbrace \phiFu;p\right\rbrace=0,
\left[(2 k \Ksept + i \Ksix) \frac{8}{k^3}\eff_2^2 - \left\lbrace \phiFu;p\right\rbrace\right]_t=0,
}
\end{array}
\right.
\end{eqnarray}
cr\'eant, outre $\Kcinq=0$, la contrainte suppl\'ementaire
\begin{eqnarray}
& & {\hskip -15.0truemm}
(k \Ksept + i \Ksix)(2 k \Ksept + i \Ksix)=0,
\end{eqnarray}
donc une seule contrainte parmi les trois param\`etres de $\CIII$,
\begin{eqnarray}
& & {\hskip -15.0truemm}
(d_2-2 d_3) (d_2-d_3)=0.
\end{eqnarray}


\textbf{Syst\`eme r\'eduit (\ref{sysUVzer}) ($f_1=0$)}. 

Le wronskien ne s'annule que pour $h_0'(t)=0$.

Si $h_0'\not=0$ et $\Kcinq\not=0$, 
les variables de r\'eduction
d\'ependent de deux fonctions arbitraires de $t$,
\begin{eqnarray}
& & {\hskip -15.0truemm}
\left\lbrace
\begin{array}{ll}
\displaystyle{
%
%
\xi= x h_0(t)+\frac{y}{h_0(t)} + h_1(t), h_0'(t)\not=0,
}\\ \displaystyle{
\du= 2 \Kcinq \left(x h_0(t)-\frac{y}{h_0(t)} \right)^2
+ 4 \Kcinq \frac{h_1'}{(\log h_0)'}\left(x h_0(t)-\frac{y}{h_0(t)} \right)\ccomma
}\\ \displaystyle{
\dv=\frac{\Ksix}{\Kcinq} (\log h_0)' \left(x h_0(t)-\frac{y}{h_0(t)} \right)\ccomma
}
\end{array}
\right.
\end{eqnarray}
et la septi\`eme \'equation (\ref{eqsyszdudv})${}_7$ d\'etermine $h_1(t)$,
\begin{eqnarray}
& & {\hskip -15.0truemm}
h_1= C_1 h_0 + \frac{C_2}{h_0} + \frac{\Ksept}{\Kcinq^2}\ccomma C_1, C_2 \hbox{ constantes arbitraires},
\end{eqnarray}
ce qui laisse donc arbitraires toutes les constantes.
\medskip

Si $h_0'\not=0$ et $\Kcinq=0$, on obtient
\begin{eqnarray}
& & {\hskip -15.0truemm}
\Ksix=0, \du=0,
\dv=\frac{4}{3} \Ksept (\log h_0)' p^3 + 4 \Ksept h_3' p^2 + \farbone(t) p,
p=x h_0(t)-\frac{y}{h_0(t)}.
\end{eqnarray}
\medskip

Si $h_0'=0$, les variables de r\'eduction
\begin{eqnarray}
& & {\hskip -15.0truemm}
\xi= x+y+t,
\du=\Fu(p,t),
\dv=\Fvt(p,t),
p=x-y,
\end{eqnarray}
d\'efinissent le syst\`eme,
\begin{eqnarray}
& & {\hskip -15.0truemm}
\Fu_{ppt} + 2 \Fu_p \Fvt_p     - 8 \Ksix =0,
- 2 \Fu_p \Fu_{pt} + \Fvt_{pp} - 8 \Ksept=0.
\label{eqsyszerFuFvt}
\end{eqnarray}
identique \`a (\ref{eqsysexpFuFvt}).

\textit{Remarque.}
Le r\'esultat (\ref{eqsysexpFuFvt}) ou (\ref{eqsyszerFuFvt})
a deux interpr\'etations.
C'est
ou bien une r\'eduction de $(u,v)(x,y,t)$ \`a une EDO $(U,V)(\xi)$
\`a coefficients $(\du,\dv)(x,y,t)$,
ou bien une r\'eduction de $(u,v)(x,y,t)$ \`a une EDP $(\Fu,\Fvt)(p,t)$
\`a coefficients $(U,V)(\xi(x,y,t))$.

\section{Conclusion} 
																						
Ce syst\`eme (\ref{eqsysuv}) tr\`es simple
admet des r\'eductions \`a cinq des six \'equations de Painlev\'e 
et \`a l'\'equation ma\^\i tresse $\CVI$ de Chazy,
sans aucune contrainte entre leurs param\`etres.

Retrouver toutes les pr\'esentes r\'eductions
par les seules m\'ethodes de la th\'eorie des groupes,
en d\'etaillant l'alg\`ebre des sym\'etries infinit\'esimales
de (\ref{eqsysuv}),
constitue un d\'efi que nous n'avons pour l'instant pas r\'eussi \`a relever.

Une discr\'etisation du pr\'esent syst\`eme (\ref{eqsysuv}) 
pourrait \^etre un bon moyen d'obtenir une version discr\`ete des \'equations $\Cn$ de Chazy.
\bigskip

\textbf{Remerciements}

Nous remercions 
chaleureusement Wolfgang Schief pour son ing\'eniosit\'e 
et
Philippe Di Francesco pour de fructueuses discussions.

Le soutien financier de 
l’Unit\'e mixte internationale UMI 3457 du Centre de recherches
math\'ematiques de l’Universit\'e de Montr\'eal 
a \'et\'e essentiel pour l'aboutissement de ces travaux.
L'auteur AMG a \'et\'e partiellement financ\'e par le CRSNG.

\vfill\eject
\appendix

\section{Les \'equations $\Cn$ de Chazy}
\label{sectionCn}

Chaque \'equation de Painlev\'e pour $u(x)$, sauf la premi\`ere,
est la cons\'equence diff\'eren\-tielle d'une \'equation de Riccati $R(u',u,x)=0$,
au moins pour une certaine relation entre les param\`etres de la $\Pn$.
Apr\`es une normalisation ad\'equate, 
le membre de gauche de ces \'equations de Riccati ob\'eit \`a de remarquables 
\'equations d'ordre deux et de degr\'e deux, signal\'ees par Chazy \cite[page 342]{ChazyThese}.

Afin de rappeler leur lien avec les $\Pn$,
nous changeons ici la notation
 (C,V),  (C,IV), (C,II), (C,III), (C,I)  de Chazy en, respectivement, 
$\CVI$, $\CVa$, $\CVb$, $\CIII$, $\CIV$,
\begin{eqnarray}
& & {\hskip -12.0truemm}
\left\lbrace
\begin{array}{ll}
\displaystyle{
\CVI: 
\left(\frac{\D^2 u}{\D x^2} -2 u^3 - d_2 u - d_3\right)^2
}\\ \displaystyle{\phantom{12}
- \left\lbrack 2 \ \fVI(x) \left(u-\frac{d_1}{\gVI(x)}\right)\right\rbrack^2
\left\lbrack \left(\frac{\D u}{\D x}\right)^2 - u^4 - d_2 u^2 -2 d_3 u - d_4\right\rbrack =0,\
}\\ \displaystyle{
\CVa: 
\left(\frac{\D^2 u}{\D x^2} -6 u^2 - d_2 u - d_3\right)^2
}\\ \displaystyle{\phantom{12}
- \left\lbrack \frac{2}{x} \left(u-\frac{x^2}{2}\right)\right\rbrack^2
\left\lbrack \left(\frac{\D u}{\D x}\right)^2 - 4 u^3 - d_2 u^2 -2 d_3 u - d_4\right\rbrack =0,\
}\\ \displaystyle{
\CVb: 
\left(\frac{\D^2 u}{\D x^2} -2 u^3 - d_2 u - d_3\right)^2
}\\ \displaystyle{\phantom{12}
+ \left\lbrack 2 \left(u-e^x\right)\right\rbrack^2
\left\lbrack \left(\frac{\D u}{\D x}\right)^2 - u^4 - d_2 u^2 -2 d_3 u - d_4\right\rbrack =0,\
}\\ \displaystyle{
\CIII: 
\left(\frac{\D^2 u}{\D x^2} - d_2 u - d_3\right)^2
- \left\lbrack 2 \frac{u}{x}\right\rbrack^2
\left\lbrack \left(\frac{\D u}{\D x}\right)^2 - d_2 u^2 -2 d_3 u - d_4\right\rbrack =0,\
}\\ \displaystyle{
\CIV: 
\left(\frac{\D^2 u}{\D x^2} -6 u^2 - d_3\right)^2
- x^2
\left\lbrack \left(\frac{\D u}{\D x}\right)^2 - 4 u^3 -2 d_3 u - d_4\right\rbrack =0,
}
\end{array}
\right.
\label{eqCn}
\end{eqnarray}
o\`u le couple $(\fVI,\gVI)$ de $\CVI$ 
est une solution quelconque du syst\`eme
\begin{eqnarray}
& & 
\left(\frac{\D f}{\D x}\right)^2=(f^2+1)^2,
\left(\frac{\D g}{\D x}\right)^2=1-g^2,
(f^2 +1) (g^2 - 1) +1=0,
\end{eqnarray}
par exemple 
$(\tg(x),\sin(x))$ (le choix de Chazy)
ou $(i \coth(x),\cosh(x))$ comme dans la solution (\ref{eqsysUVtrisol}).

Leur int\'egrale g\'en\'erale \cite[Appendix]{Cos2006b} est le produit de $R(u',u,x)$ par un polyn\^ome de $u$ et $x$.

\section{Variable r\'eduite g\'en\'erique}
\label{section-xi}
\'Etant donn\'e l'\'equation (\ref{pdeab}),
l'\'elimination de $\fb(y,t)$ (resp.~$\fa(x,t)$) engendre deux d\'eriv\'ees d'EDOs,
\begin{eqnarray}
& &  {\hskip -10.0 truemm} 
(\lbrace \fa(x,t);x\rbrace)_t=0, 
(\lbrace \fb(y,t);y\rbrace)_t=0,
\label{odea-odeb} 
\end{eqnarray}
o\`u la notation classique $\lbrace f;x\rbrace$ d\'esigne le schwarzien 
\begin{eqnarray}
& &  {\hskip -10.0 truemm} 
\lbrace f;x \rbrace
=\frac{f_{xxx}}{f_x} - \frac{3}{2} \left(\frac{f_{xx}} {f_x} \right)^2.
\end{eqnarray}
La solution g\'en\'erale de chacune des deux \'equations (\ref{odea-odeb})
d\'epend de quatre fonctions arbitraires d'une variable
%
\begin{eqnarray}
& &  {\hskip -10.0 truemm} 
\fa(x,t)=\eff_1(t)+\frac{\eff_2(t)}{\fx(x)-\eff_3(t)}\ccomma
\fb(y,t)=\egg_1(t)+\frac{\egg_2(t)}{\fy(y)-\egg_3(t)}\ccomma
\label{eqsolpdeab}
\end{eqnarray} 
li\'ees par trois contraintes
%
\begin{eqnarray}
& &  {\hskip -10.0 truemm} 
\egg_1=-\eff_1, \egg_2=-\eff_2, \egg_3=\eff_3,
\label{eqsolpdeab-contraintes}
\end{eqnarray} 
ce qui laisse donc cinq fonctions arbitraires.
L'invariance conforme permet le choix $\fx(x)=x,\fy(y)=y$.


\vfill \eject
\end{document}